\documentclass[]{aa}  
\usepackage{graphicx}
\usepackage{txfonts}
\usepackage[]{hyperref}

\begin{document} 

   \title{GW190814 follow-up with the optical telescope MeerLICHT}

   \author{S. de Wet \inst{1,}\thanks{Corresponding author \email{DWTSIM002@myuct.ac.za} }
          \and
         P. J. Groot \inst{1,}\inst{2,}\inst{3}
         \and
         S. Bloemen\inst{2}
         \and
         R. Le Poole\inst{4}
         \and
         M. Klein-Wolt\inst{2}
         \and
         E. K\"ording\inst{2}
         \and
         V. McBride\inst{1,5}
         \and
         K. Paterson\inst{6}
         \and
         D. L. A. Pieterse\inst{2}
         \and
         P. M. Vreeswijk\inst{2}
         \and
         P. Woudt\inst{1}
          }

   \institute{Inter-University Institute for Data Intensive Astronomy \& Department of Astronomy, University of Cape Town, Private Bag X3, Rondebosch, 7701, South Africa
              \and
              Department of Astrophysics/IMAPP, Radboud University, P.O. Box 9010, 6500 GL, Nijmegen, The Netherlands
              \and 
              South African Astronomical Observatory, P.O. Box 9, 7935, Observatory, South Africa
              \and
              Leiden Observatory, Leiden University, P.O. Box 9513, NL-2300 RA Leiden, The Netherlands
              \and
              IAU-Office For Astronomy for Development, P.O. Box 9, 7935, Observatory, South Africa
              \and
              Center for Interdisciplinary Exploration and Research in Astrophysics (CIERA), Northwestern University, 1800 Sherman Ave, Evanston, IL 60201, USA
             }

   \date{Received \today; accepted \today}

  \abstract
   {The Advanced LIGO and Virgo gravitational wave observatories detected a signal on 2019 August 14 during their third observing run, named GW190814. A large number of electromagnetic facilities conducted follow-up campaigns in the search for a possible counterpart to the gravitational wave event, which was made especially promising given the early source classification of a neutron star-black hole merger. }
   {We present the results of the GW follow-up campaign taken with the wide-field optical telescope MeerLICHT, located at the South African Astronomical Observatory Sutherland site. We use our results to constrain possible kilonova models. }
   {MeerLICHT observed more than 95\% of the probability localisation each night for over a week in three optical bands (\textit{u,q,i}) with our initial observations beginning almost 2 hours after the GW detection. We describe the search for new transients in MeerLICHT data and investigate how our limiting magnitudes can be used to constrain an AT2017gfo-like kilonova. }
   {A single new transient was found in our analysis of MeerLICHT data, which we exclude from being the electromagnetic counterpart to GW190814 due to the existence of a spatially unresolved source at the transient's coordinates in archival data. Using our limiting magnitudes, the confidence with which we can exclude the presence of an AT2017gfo-like kilonova at the distance of GW190814 was low ($<10^{-4}$). }
   {}


   \maketitle

\section{Introduction}

The detection of the first binary black hole merger (BBH) in GW150914 \citep{GW150914} opened up the era of gravitational wave astronomy, with a further 9 confirmed BBH mergers detected during the first two observing runs (O1 and O2) of the LIGO Scientific and Virgo Collaboration (LVC), along with an additional 3 BBH candidates found through independent analysis \citep{zackay1,zackay2}. The first -- and currently only -- multi-messenger source was detected during O2 and was caused by the merger of two neutron stars in a binary system (BNS) \citep{GW170817a,GW170817b}. The electromagnetic (EM) counterparts to GW170817 were observed across the EM spectrum by numerous observing facilities \citep{GW170817c,goldstein,savchenko,coulter,lipunov,tanvir,soares,valenti} with implications across a vast range of scientific disciplines. Optical/near-infrared observations demonstrated that the emission was due to a kilonova (KN) \citep{arcavi,chornock,covino,cowperthwaite,drout,evans,kasliwal,mccully,nicholl,pian,shappee,smartt,tanvir} powered by the radioactive decay of r-process material produced during the merger \citep{metzger,kasen,gall,watson2019}. For the first time short gamma-ray bursts were convincingly linked to BNS mergers due to a coincidental detection of a gamma-ray signal \citep{lyman,dobie,mooley,lazzati,resmi,ghirlanda,lamb,margutti,nynka,troja,davanzo}.

The third LVC observing run (O3) began 2019 April 1 and concluded 2020 March 27 with a total of 39 candidate events detected over the first half of the run (O3a) -- a major increase from the 11 events detected over the course of O1 and O2 \citep{O3a}. A number of scientifically rich discoveries have come out of O3: the event GW190412 revealed the first BBH merger with a clearly unequal mass ratio along with significant higher-multipole gravitational radiation \citep{GW190412}; GW190425 \citep{GW190425} was the second binary neutron star merger detected in gravitational waves; GW190521 was produced by the most massive BBH system yet detected \citep{GW190521}; and GW190814 was the result of a compact binary coalescence with the most unequal mass ratio yet measured in gravitational waves, with the secondary component having a mass that would make it either the lightest BH or heaviest NS ever discovered \citep{GW190814}. 

\subsection{GW190814}
 A preliminary GCN Notice sent out by the LVC at 21:31:40 UT on 2019 August 14 indicated that a gravitational wave event had been detected in data from LIGO Livingston and Virgo at 21:11:00 UTC. The event was given the superevent ID S190814bv. The 90\% probability region had an area of 38 deg$^2$ at a luminosity distance of $276\pm56$ Mpc, with an extremely low false alarm rate of one event per $1.559\times10^{25}$ years \citep{GCN25324}. The early classification as a neutron star black hole (NSBH) merger along with its small sky
 localisation and low false alarm rate made it a promising candidate for EM follow-up. Campaigns were undertaken by numerous EM facilities and neutrino facilities \citep{dobie2019,gomez,lipunovgcn,ackley,antier,andreoni,watson,vieira,ageron,icecube}, with no viable counterpart being found. 

Further analysis of the GW190814 data \citep{GW190814} revealed that the 90\% probability region encompassed 18.5 deg$^2$ at a distance of $241^{+41}_{-45}$ Mpc and was caused by a compact binary coalescence with a mass ratio $q=m_2/m_1=0.112$. The secondary component of the binary had a mass of 2.59M$_{\odot}$, making it either the lightest BH or heaviest NS yet discovered. The primary component was classified as a BH with a mass of 23.2M$_{\odot}$ and dimensionless spin tightly constrained to $\chi_1\leq0.07$. The lack of any EM counterpart is in agreement with their assessment that the secondary component was unlikely to have been a NS based on existing estimates of the maximum NS mass and was therefore likely caused by a BBH merger, an assessment supported by further studies \citep{essick,tews}. This novel discovery has challenged population synthesis models and existing assumptions on the lightest BH or heaviest NS \citep{GW190814}. 

In addition to GW190814, the EM community was active in its follow-up of O3 events, particularly events caused by binaries that likely contained at least one NS. Ten such events occurred during O3a \citep{coughlin_a} and a further 5 in the remainder of O3 \citep{coughlin_b}. Despite the large increase in candidates for EM follow-up compared to O1 and O2, no significant counterparts to any of these events were detected, owing in part to their large distances and sky-areas. Strong limits were placed on any counterparts to the BNS candidate GW190425 by a number of groups \citep{coughlin_c,hosseinzadeh}, and a counterpart to GW190521 was reported, making it the first BBH with a strong candidate counterpart \citep{graham}.  

The focus of this paper is GW190814 and the follow-up campaign conducted by the MeerLICHT optical telescope in Sutherland, which observed more than 95\% of the probability localisation each night for over a week in three optical bands (\textit{u, q} and \textit{i}), with our initial observations being some of the earliest optical data taken by any group -- the first observation beginning almost 2 hours after the GW detection. In \S \ref{sec:ML} we introduce the MeerLICHT optical telescope and the GW190814 follow-up observing campaign taken with that telescope. In \S \ref{sec:transients} the search for transients in MeerLICHT data is described, and in \S \ref{sec:results} we show how our limiting magnitudes were used to constrain the possible KN parameter space. All magnitudes - unless stated otherwise - are given in the AB magnitude system.

\section{Observations with MeerLICHT}\label{sec:ML}
MeerLICHT is a wide-field and fully robotic optical telescope situated at the South African Astronomical Observatory site near Sutherland. Designed and built as a prototype for the BlackGEM array \citep{groot}, its primary science goals centre around its novel pairing with the 64-antennae MeerKAT radio array where it will provide simultaneous night-time, multi-filter optical coverage of the radio sky as observed by MeerKAT. The telescope possesses a 65 cm primary mirror with a 110 Megapixel CCD resulting in a  2.7 deg$^2$ field of view sampled at 0.56"/pixel \citep{bloemen}. The 6-filter wheel consisting of 5 SDSS filters (\textit{u,g,r,i,z}) plus the wider \textit{q} band (440-720 nm) make it perfectly suited to the multi-colour study of the transient sky. 

MeerLICHT's general strategy for GW follow-up is to cover the full sky localisation every 2 hours. If the sky area is small enough, it will be covered in the \textit{u, q} and \textit{i} bands as they collectively encompass most of the optical portion of the EM spectrum, allowing us to probe colour evolution independent of KN models. If the sky area is large, a single band will be used, usually \textit{q}. The choice of exposure time involves balancing the benefits of deeper limits gained from longer exposures with the associated reduction in sky coverage. Since MeerLICHT is sky background limited for 60-second exposures, we employ such an exposure time for GW follow-up with the option of performing co-addition of exposures to achieve deeper limits during post-analysis. GW190814 follow-up observations were initially planned using the second \texttt{BAYESTAR} sky map made publicly available on the \href{https://gracedb.ligo.org/superevents/S190814bv/view/}{GraceDB} website at 22:58:20 UTC on 2019 August 14. The ranked-tiling method \citep{ghosh} was used to determine which of MeerLICHT's fixed sky-grid fields should be observed to cumulatively encompass a probability of at least 95\%. A total of 24 such fields were identified. Over the course of the first night a total of 191 exposures of 60-second duration were taken in the \textit{u, q} and \textit{i} bands. Our first observations began 2 hours after the GW detection at 23:11:33 UTC, and by 01:06:30 on August 15 (3.92 hours post-detection) we had observed all 24 fields at least once in each of our 3 bands, making us one of the earliest groups to observe the GW localisation in its entirety. 

\begin{figure}
   \centering
   \includegraphics[width=\hsize]{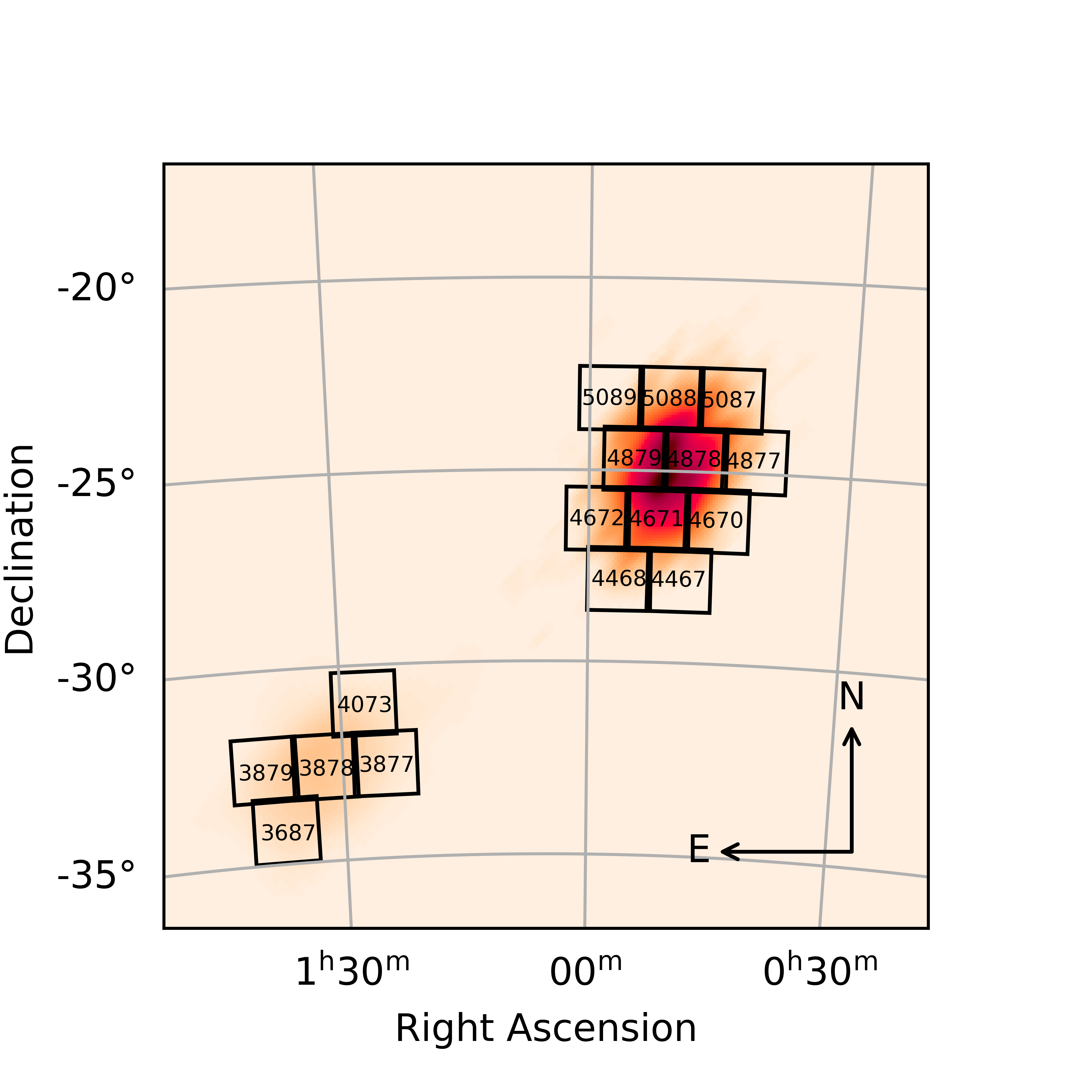}
      \caption{Sky positions of the 16 MeerLICHT fields that encompassed a probability of at least 95\% of the \texttt{\texttt{LALInference}} sky map. The numbers in each tile denote that field's unique Field ID. See Table \ref{tab:limmags} for the coordinates and integrated probability of each field.}
         \label{fig:skymap}
   \end{figure}

The \texttt{LALInference} sky map released the following day reduced the 90\% probability region from 38 to 23 deg$^2$. Implementation of the ranked-tiling method reduced the required number of fields from 24 to 16. For the remainder of the observing time spent on the follow-up of GW190814, the 16 fields shown in Fig. \ref{fig:skymap} were observed each night. A total of 1484 exposures were taken between August 14 and 24. 

At the time of GW190814, the MeerLICHT observing system was still in the commissioning phase, with troubleshooting taking place on an ongoing basis. Analysis of this dataset alerted the MeerLICHT team to a problem with the rotation of the telescope dome while observing. Unfortunately, a large fraction of the data could not be used ($\sim56\%$) due to vignetting caused by the telescope's dome. Subsequent scientific analysis was therefore undertaken using the unaffected data. Table \ref{tab:limmags} presents all the observations used in this paper.

\section{Search for transients in MeerLICHT data}\label{sec:transients}
The software pipeline for reducing MeerLICHT's raw images was initially largely based on SkyMapper's \citep{skymapper} with modifications and additions, but now stands independently. Written in Python the software consists of two components: the first is known as BlackBOX\footnote{Source code available at \href{https://github.com/pmvreeswijk/BlackBOX}{https://github.com/pmvreeswijk/BlackBOX}.} which performs standard CCD reduction tasks on the raw science images; the second is ZOGY\footnote{Source code available at \href{https://github.com/pmvreeswijk/ZOGY}{https://github.com/pmvreeswijk/ZOGY}. } which is used for identifying sources, performing astrometry and photometry, and finding transients through the optimal image subtraction routines formulated by \cite{zogy}. The method uses statistical principles to derive the optimal statistic for transient detection, taking into account the point spread functions of both the new and reference images to produce the difference image. The significance image contains the probability of a transient being present at a particular location or pixel, while the corrected significance (\textit{S}\textsubscript{corr}) image normalises the significance image using the source and background noise and astrometric uncertainties, resulting in an image having units of sigmas in which errors due to bad subtractions are less likely to show up. Candidate transients are identified from the \textit{S}\textsubscript{corr} images. All sources having a signal-to-noise ratio $|\text{S/N}|\geq6$ are included in a transient catalogue file associated with the new image. A positive \textit{S}\textsubscript{corr} value for a source indicates that the source is new or has brightened with respect to the reference image, while negative values indicate that the source has faded. Since no MeerLICHT data of the 16 GW follow-up fields existed prior to our observations, deep reference images were created by co-adding all images in a particular filter taken over the course of the follow-up campaign. This did have the drawback that any persistent transient would also be present in the reference image and hence only significantly fast-evolving sources over the course of the observations would be flagged as transient candidates.

\begin{figure}
   \centering
   \includegraphics[width=\hsize]{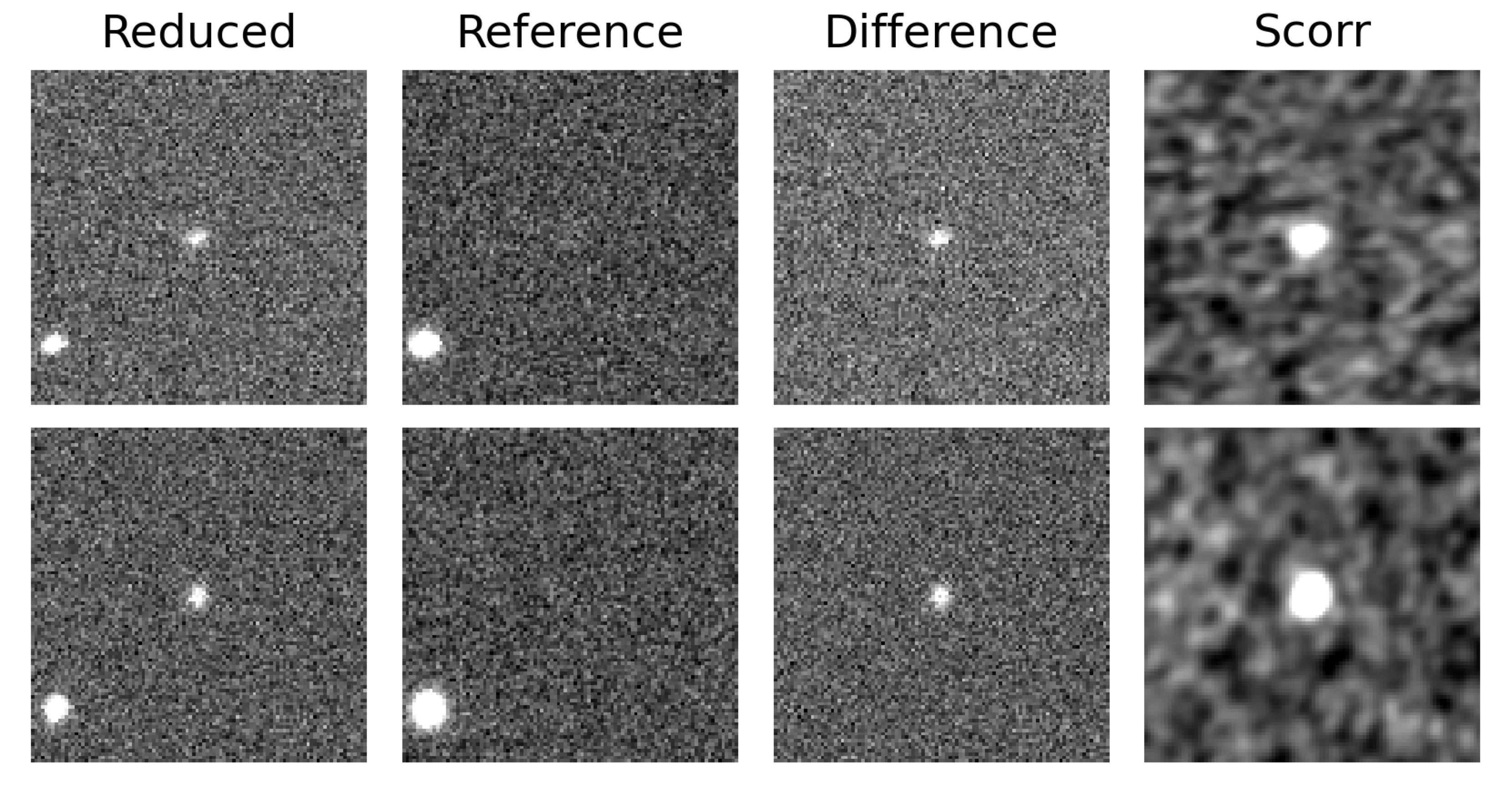}
      \caption{Vetting cutouts for the matching pair identified at coordinates ($23.44258^{\circ},-32.67500^{\circ}$) in field 3878 on the night of 2019 August 20. The top row of cutouts corresponds to an \textit{i}-band detection, while the bottom row corresponds to a \textit{q}-band detection. The time interval between both observations was approximately 1.5 minutes. The source was identified in the Minor Planet Center database of known objects. }
         \label{fig:matching_pair}
   \end{figure}

The search for transients in our GW follow-up data was undertaken using the transient catalogue files. To reduce the number of potential bogus candidates, we added the constraint that any source must have had at least two transient detections on a particular night, regardless of the filter of the observations. To do this, a list of all unique combinations of transient file pairs from a particular field and night was created. A crossmatch was performed across each pair of files using a $1\arcsec$ sky radius. Pairs of sources occurring in both files -- known hereon as a matching pair -- would be manually vetted. Across all 16 fields, 545 matching pairs were identified. These were manually vetted by comparing $1'\times1'$ cutouts of the reduced, reference, difference, and \textit{S}\textsubscript{corr} images centred on the coordinates of the source, as shown in Fig. \ref{fig:matching_pair}. 

A total of 455 candidates were identified after removal of bogus candidates, of which 43 did not have a clear source present in their reference images. The remaining 412 sources were likely variable or flaring stars. The \href{https://gea.esac.esa.int/archive/}{Gaia DR2} \citep{gaia} database was queried to determine which of these 412 sources had detections in the Gaia database. All but one of the sources had a match within $3\arcsec$ of the MeerLICHT coordinates. The single source without a Gaia detection was found at the core of the Seyfert II galaxy \href{http://simbad.u-strasbg.fr/simbad/sim-id?Ident=ESO+353-9&NbIdent=1&Radius=2&Radius.unit=arcmin&submit=submit+id}{ESO 353$-$9}, which we exclude from being the counterpart to GW190814 due to its lower redshift of $z=0.0167$ \citep{eso353} compared to $z=0.053^{+0.009}_{-0.010}$ for GW190814 \citep{GW190814}. We suspect that the core of the galaxy was showing variable behaviour, which could explain its detection as a transient candidate. For the 411 sources with Gaia matches, the \texttt{CLASS\_STAR}\footnote{The \texttt{CLASS\_STAR} star/galaxy classifier is implemented by SExtractor \citep{sextractor} during reduction.} catalogue parameter in the MeerLICHT catalogue files was used to exclude them from being transients in a respective host galaxy. A \texttt{CLASS\_STAR}  value close to 1 indicates that a source is likely stellar, while a value close to 0 indicates that the source is likely a galaxy. For each source, the mean \texttt{CLASS\_STAR} value for all catalogue entries in a particular filter was calculated. Sources with at least one mean \texttt{CLASS\_STAR} value greater than 0.5 were regarded as stellar. A single source did not meet this requirement: the quasar \href{http://simbad.u-strasbg.fr/simbad/sim-id?Ident=\%401278296&Name=QSO\%20B0035-252&submit=submit}{QSO B0035-252} had a mean \texttt{CLASS\_STAR} value less than 0.06 in all filters. The source had been reported to the TNS on 2019 August 19 by the MASTER group \citep{MASTER} and given name AT2019nvx, though from the light curve it simply appeared to be showing variable behaviour. We therefore exclude all 412 sources with Gaia matches from being associated with GW190814. 

The 43 matching pairs without any clear source in their reference images were checked for being asteroids. Based on the orbits of known objects in the \href{https://www.minorplanetcenter.net/data}{Minor Planet Center (MPC)} database, 27 of the 43 candidates were positively identified as asteroids. For the remaining candidates, light curves were produced by searching for $1\arcsec$ matches at the candidates' coordinates in the catalogue files associated with each reduced image. Any candidate with only two detections in its light curve -- corresponding to the two detections making up the matching pair -- would likely have been an asteroid. The asteroid status of each candidate was confirmed by visually examining the images of a particular field from the night of the transient detection for movement of the source across the field. All but one of the sources were confirmed as asteroids in this way. 

\subsection{Astrophysical transient candidate MLT J012825.10$-$312414.4}
\begin{figure}
   \centering
   \includegraphics[width=\hsize]{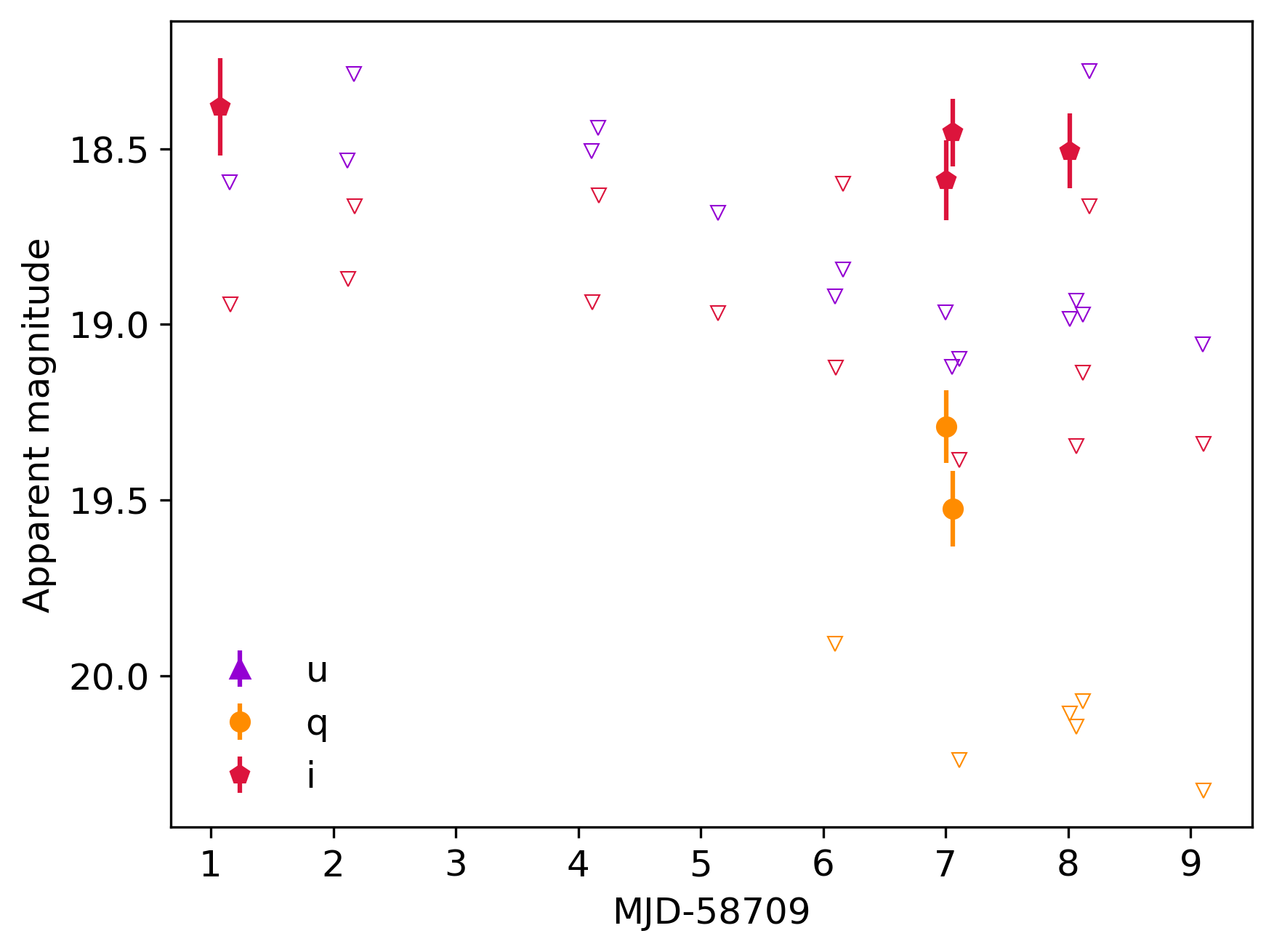}
      \caption{Light curve of MLT J012825.10-312414.4 located in field 4073. The source was detected at 6 epochs on 3 separate nights in the \textit{q} and \textit{i} bands. Upside-down triangles indicate $5\sigma$ limiting magnitudes. }
         \label{fig:light_curve}
   \end{figure}

The remaining candidate was detected on 3 separate nights and is astrophysical in origin; its light curve is shown in Fig. \ref{fig:light_curve}. No source was found at these coordinates in the \href{https://wis-tns.weizmann.ac.il}{Transient Name Server} database, hence we regard MLT J012825.10$-$312414.4 as a new transient candidate. A search of the \href{http://vizier.u-strasbg.fr}{VizieR} Catalogue Service revealed a number of survey detections near (all $\lesssim1.7$") the transient's coordinates. The Pan-STARRS PS1 catalogue \citep{panstarrs} had a point source detection 1.7" from the transient's coordinates with \textit{i}-band magnitude $i=21.43\pm0.03$. We associated this source with MLT J012825.10$-$312414.4 through comparison of the MeerLICHT and Pan-STARRS images. We note that the MeerLICHT detections of MLT J012825.10$-$312414.4 were all near the detection threshold of the telescope, which increased the uncertainty in position. Source detections in the Dark Energy Survey \citep{DES} and AllWISE \citep{AllWISE} catalogs are also likely to be associated with MLT J012825.10$-$312414.4. The existence of these associated source detections in archival data means that we can rule out MLT J012825.10$-$312414.4 as the EM counterpart to GW190814.

\section{Constraints on KN parameter space}\label{sec:results}

The lack of any viable electromagnetic counterpart to GW190814 in MeerLICHT data was in agreement with the findings of other groups, and also expected since the probability of detectable EM emission was low given that GW190814 was produced by either a BBH or high mass-ratio NSBH \citep{GW190814}. Nevertheless, we use our limiting magnitudes to place limits on any potential counterpart and constrain possible KN models. 

\subsection{Comparison with AT2017gfo-like KN} \label{sec:compare_at2017gfo}

\begin{figure}
   \centering
   \includegraphics[width=\hsize]{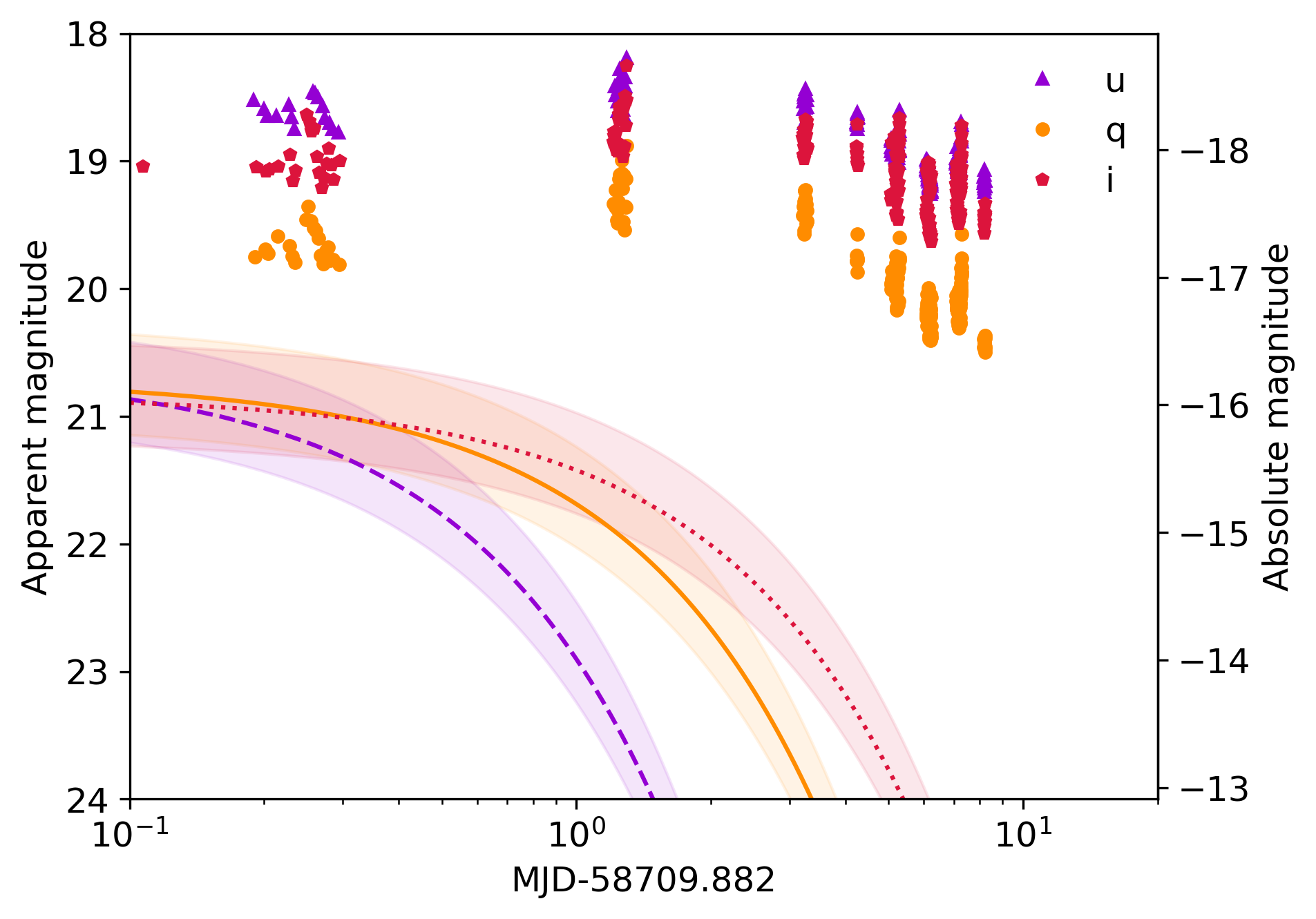}
      \caption{Phenomenological fits to the AT2017gfo light curve (with \textit{u} dashed, \textit{q} smooth and \textit{i} dotted) are shown in relation to MeerLICHT's $5\sigma$ limiting magnitudes. The shaded regions indicate the 1$\sigma$ distance uncertainties. Apparent magnitudes were converted to absolute magnitudes using $M=m-5\textrm{log}_{10}(d_L/10\textrm{pc})$ at the distance of GW190814.}
         \label{fig:compare}
   \end{figure}

It is instructive to compare the limiting magnitudes from our observations with the light curve of the only confirmed kilonova to date -- AT2017gfo. We followed the approach of \cite{ackley} by performing phenomenological fits to the AT2017gfo light curve in each of the relevant bands so as to compare our limits with a possible AT2017gfo-like KN. Photometric data on AT2017gfo was obtained from the compilation of light curves associated with that event in \cite{Villar}. Combined \textit{U}- and \textit{u}-band data was used for our \textit{u}-band fit; \textit{V}-band data was regarded as a proxy for MeerLICHT's \textit{q}-band as they both have similar central wavelengths, though our \textit{q}-band is wider; and the \textit{i}-band data was naturally used for our \textit{i}-band. \cite{Gompertz} fit either an exponential or Bazin function depending on if a clear peak is visible in the light curve post-merger. Exponential curves were fitted to each of the three light curves in flux space and converted back to magnitudes. The model fits for each band were then converted to apparent magnitudes at the distance of GW190814. Using distances of 40 Mpc and $241^{+41}_{-45}$ Mpc to AT2017gfo and GW190814 respectively, the conversion amounted to a shift of 4.12 magnitudes. Our limiting magnitudes from each 60-second exposure are shown in relation to these model fits in Fig. \ref{fig:compare}. Our most sensitive limits are in the wider \textit{q} band, though even the deepest limits are insensitive to an AT2017gfo-like KN by at least a magnitude. 

\begin{figure}
   \centering
   \includegraphics[width=\hsize]{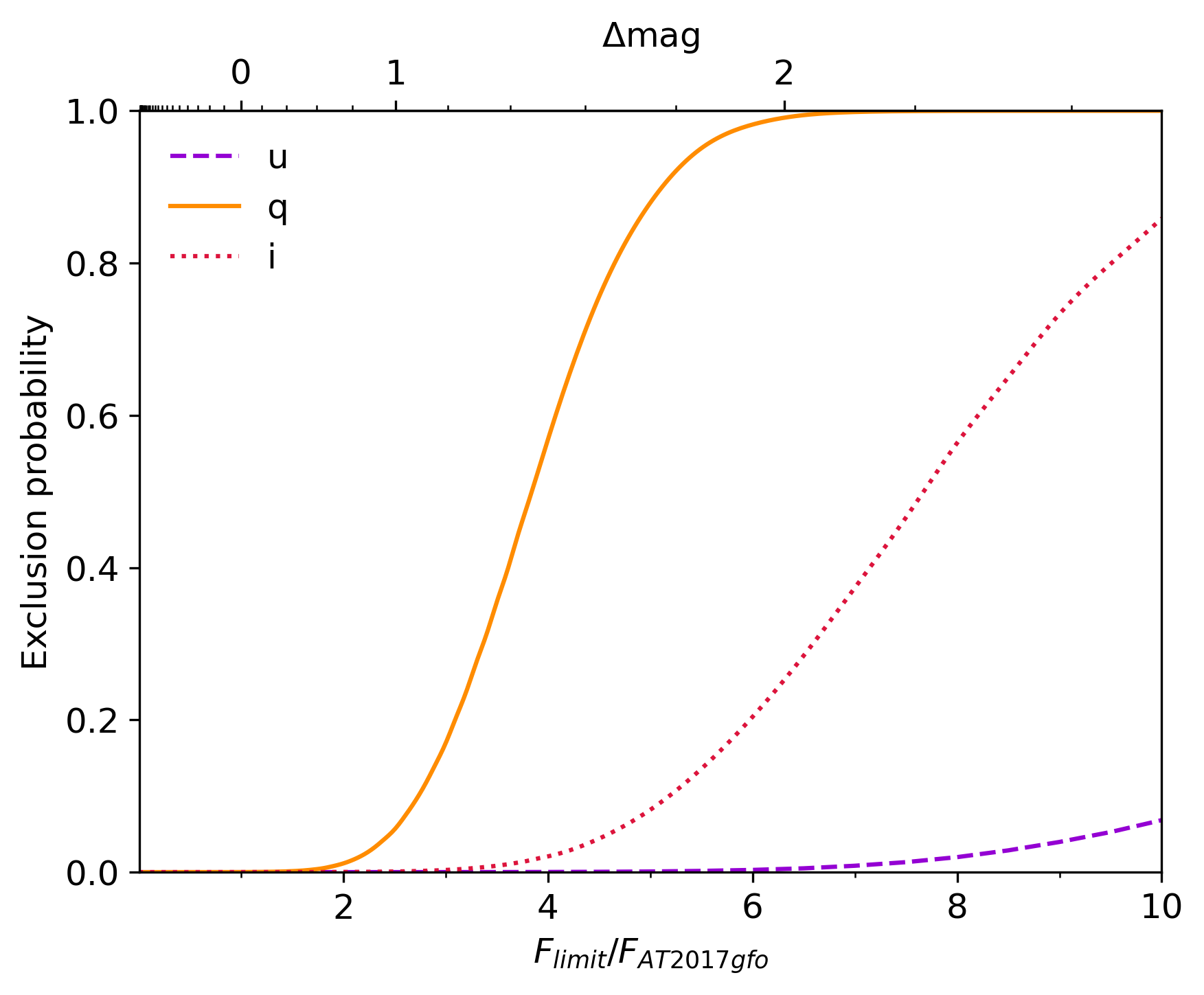}
      \caption{The exclusion probability of our observations as a function of scaled AT2017gfo-like KN flux (bottom axis) and corresponding vertical shift of the light curves in Fig. \ref{fig:compare} in magnitudes (top axis). An $x$-value of 1 corresponds to an AT2017gfo KN at the distance of GW190814. }
         \label{fig:covered_probability}
   \end{figure}

The exclusion probability is a measure of the confidence with which we can exclude an EM counterpart model given our wide-field observations and the GW 3D skymap. We can calculate the exclusion probability of our observations assuming an AT2017gfo-like KN by using our limiting magnitudes and model light curves combined with the 3D sky probability distribution for GW190814. We followed the approach for a wide-field search as outlined in Appendix A of \cite{ackley}, making use of their equations A.4 and A.5. We scaled the AT2017gfo model flux by a range of multiplicative factors between 0 and 10. These have the effect of shifting the Fig. \ref{fig:compare} light curves up (for factors $>1$) or down (for factors $<1$). Figure \ref{fig:covered_probability} demonstrates that the exclusion probability of our observations assuming an AT2017gfo-like KN is very low -- in \textit{q} it is only $8.75\times 10^{-5}$, and even lower in \textit{u} and \textit{i}. In our most sensitive band -- \textit{q} -- the exclusion probability is close to unity for a KN 5 times brighter than AT2017gfo. 

\subsection{Exclusion probability in mass ejecta-viewing angle parameter space}

A number of KN light curve models are available which can be used to constrain the physical parameters of a possible KN using observational data. Popular models include those of  \cite{kasen}, \cite{Bulla}, and \cite{hotokezaka}, as used in the multi-model analyses of \cite{dietrich} and \cite{coughlin_a, coughlin_b}. Each model depends on a number of physical properties: the model of \cite{kasen} depends on the ejecta mass, mass fraction of lanthanides, and ejecta velocity; the 2-component semi-analytic model of  \cite{hotokezaka} depends on the ejecta mass, ejecta velocity, the dividing velocity between the inner and outer part, and the opacity of the 2 components.

We perform our analysis with a single KN model - the time-dependent 3D Monte Carlo code POSSIS outlined in \cite{Bulla}, which depends on the total ejecta mass ($M_\text{ej}$), the half-opening angle $\Phi$, and the temperature $T$ of the ejecta at one day post-merger. The code models radiation transport in supernovae and kilonovae using wavelength and time-dependent opacities. For KNe, it assumes a spherical two component ejecta model consisting of a lanthanide-rich component distributed around the equatorial plane with half-opening angle $\Phi$, and a lanthanide-free component at higher latitudes. The lanthanide-rich component can be thought of as the dynamical ejecta, and the lanthanide-free component as the disk wind ejecta \citep{Bulla}. An advantage of this model over others is that it produces viewing-angle dependent observables. A number of these models appear in the papers of \cite{Bulla} and \cite{Dhawan}, and have been made \href{https://github.com/mbulla/kilonova_models}{publicly available}. Models are computed for ejecta mass values in the range $[0.01, 0.10]$ in steps of $0.01M_{\odot}$, $\Phi$ in $[15, 75]$ in steps of $15^{\circ}$, and $T$ in $[3000, 9000]$ in steps of 2000K. For our analysis we only considered models with temperatures $T=5000$ K as this was the best-fit value to AT2017gfo found by \cite{Dhawan}. For each model, spectral energy distributions (SEDs) in the wavelength range $0.1-2.3$ $\mu$m ($\Delta\lambda=0.022$ $\mu$m) were available at times ranging from 0.5 to 15 days post-merger in time steps of 0.5 days, for 11 viewing angles equally spaced in cos($\theta_\text{obs}$) in the range $[0,1]$. Viewing angles varied from face-on ($\textrm{cos}(\theta_\textrm{obs})=1$) to edge-on ($\textrm{cos}(\theta_\textrm{obs})=0$). Using the SEDs -- where the fluxes are given at a distance of 10 pc -- along with MeerLICHT's filter transmission curves, absolute AB magnitudes in a particular filter could be calculated by integrating the flux. Light curves can thus be easily extracted and converted to apparent magnitudes at the desired distance. 

\begin{figure*}
   \centering
   \includegraphics[width=0.9\hsize]{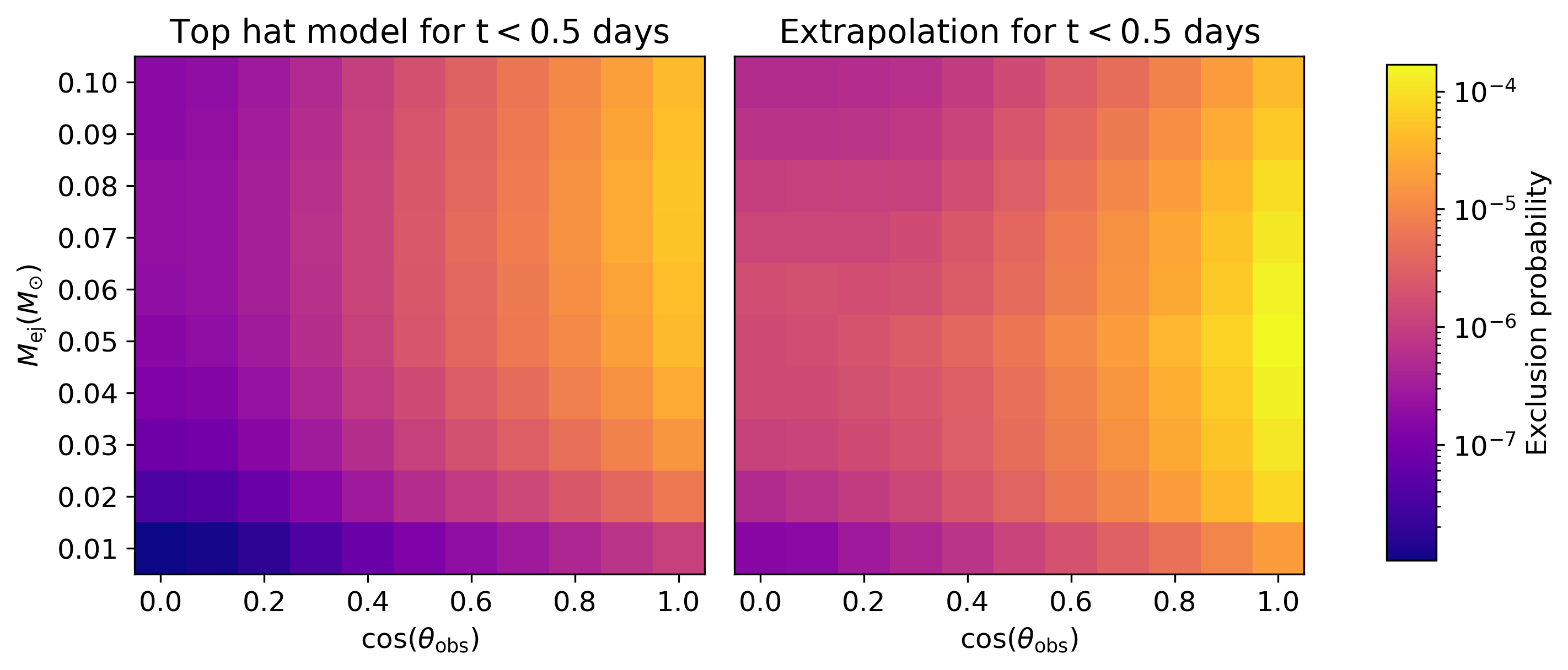}
      \caption{Exclusion probability of our \textit{q} band observations for KN models with varying ejecta mass and viewing angle. The left-hand plot employs a top hat model at early times (t $<0.5$ days) post-merger, while the right-hand plot extrapolates the model light curves using a cubic spline. }
         \label{fig:possis_covprob}
   \end{figure*}

In the same way as in \S \ref{sec:compare_at2017gfo}, we can calculate the exclusion probability of our observations but instead using a POSSIS KN model. Setting $\Phi=30^{\circ}$ -- the best-fit value to AT2017gfo found by \cite{Bulla} and \cite{Dhawan} -- we calculated the exclusion probability of our observations for a variety of KN models, varying the ejecta mass and viewing angle. For times earlier than 0.5 days post-merger it is unclear how the light curves should behave, so we used two methods to compare our earliest limiting magnitudes with the model light curves: first we adopted a top hat model interpolation scheme where the light curves are held constant at the first available model value for times t $<0.5$ days; secondly we extrapolated the model light curves using a cubic spline for t $<0.5$ days. Figure \ref{fig:possis_covprob} presents the exclusion probability in the \textit{q} band for the range of KN models for both methods. For both methods the exclusion probability increases for more polar viewing angles which is probably due to the fact that more polar viewing angles result in brighter KNe. A more interesting trend is that the exclusion probability does not peak at the highest ejecta mass of $0.10M_{\odot}$ as would be expected for increasing ejecta mass \citep[see][]{Bulla}. Instead the top hat method peaks at an ejecta mass of $0.08M_{\odot}$, while the extrapolation method peaks at $0.05M_{\odot}$. This is due to the nature of the KN models, where it is not guaranteed that KNe with higher ejecta mass have brighter peaks in their light curves, even though the bolometric luminosity will be greater for larger ejecta masses. Since there is still much uncertainty surrounding the behaviour of the KN models at early (t $<0.5$ days) times post-merger, we caution for an over-interpretation of this result and encourage the calculation of earlier-time models. 

\section{Discussion}
We found one new transient -- MLT J012825.10$-$312414.4 -- in MeerLICHT data on GW190814, which we excluded from being the counterpart to GW190814 due to the existence of a spatially unresolved source in archival Pan-STARRS data. As demonstrated in Fig. \ref{fig:compare} and Fig. \ref{fig:covered_probability}, our observations were not sufficiently deep to exclude any AT2017gfo-like KN at the distance of GW190814. Based on our limiting magnitudes per field, we would likely have detected such a KN out to a distance of 56, 132 and 95 Mpc in the \textit{u}, \textit{q} and \textit{i} bands, respectively. On average, we took three 60-second exposures per field and filter each night. Nightly co-additions of these images would have allowed us to probe deeper (by $\sim0.6$ mag) and increase the exclusion probability of our observations, had there not been a significant loss of data caused by vignetting. We also note that our effective depth was adversely affected by the moon being full during the days immediately following the GW detection. The absence of any EM counterpart to GW190814 was expected in light of the high probability that the event was caused by a BBH or high mass-ratio NSBH merger.

\section{Conclusion}
The prospect of finding an EM counterpart to a high significance GW event detected by the LVC on 2019 August 14 was made particularly promising given its early classification as a NSBH merger. Numerous groups conducted follow-up observations which were facilitated by the small sky-localisation of approximately 23 deg$^2$ at the 90\% credible level. The MeerLICHT optical telescope in Sutherland observed the GW localisation each night for more than a week, covering at least 95\% of the localisation probability in three bands ($u$, $q$, and $i$), often 3 or more times per night per band. The median depth per exposure of our observations (in the AB magnitude system) was 18.98 in $u$, 20.02 in $q$, and 19.09 in $i$. We found one new transient in our analysis, which we rule out being the EM counterpart to GW190814 due to the existence of a spatially unresolved source at the transient's coordinates in archival Pan-STARRS data. We used MeerLICHT limiting magnitudes to calculate the covered probability of our observations assuming an AT2017gfo-like KN at the distance of GW190814. Our covered probability in all 3 bands was negligible ($<10^{-4}$), however it is highly probable that we would have been able to detect a KN approximately 5 times brighter than AT2017gfo, at the distance of GW190814. Furthermore, we used our limiting magnitudes to investigate the mass ejecta-viewing angle parameter space of KN models produced by the time-dependent 3D Monte Carlo code POSSIS. For KNe with a half-opening angle of $30^{\circ}$ we found that ejecta masses of $0.08M_{\odot}$ (using an early-time top hat model) and $0.05M_{\odot}$ (using early-time extrapolation) with an edge-on viewing angle had the greatest probabilities of being observed, though still very low ($p\sim10^{-4}$). 

\begin{acknowledgements}
This research has made use of data and/or services provided by the International Astronomical Union's Minor Planet Center. The MeerLICHT consortium is a partnership between Radboud University, the University of Cape Town, the Netherlands Organisation forScientific Research (NWO), the South African Astronomical Observatory (SAAO), the University of Oxford, the University of Manchester and the University of Amsterdam, in association with and, partly supported by, the South African Radio Astronomy Observatory (SARAO), the European Research Council and the Netherlands Research School for Astronomy (NOVA). We acknowledge the use of the Inter-University Institute for Data Intensive Astronomy (IDIA) data intensive research cloud for data processing. IDIA is a South African university partnership involving the University of Cape Town, the University of Pretoria and the University of the Western Cape. PAW acknowledges support from the NRF and UCT. SdW and PJG are supported by NRF SARChI Grant 111692. 
\end{acknowledgements}

\bibliographystyle{aa}
\bibliography{refs.bib}

\begin{appendix} 
\section{Usable data and limiting magnitudes}\label{app:data}
From the 1484 raw images taken over the course of our GW190814 follow-up campaign we identified 649 images that were unaffected by vignetting, as listed in Table \ref{tab:limmags}. Subsequent analysis was undertaken using this dataset.

\longtab[1]{
\begin{longtable}{lllllll}
\caption{List of usable observations taken with MeerLICHT. For each Field ID we provide the RA and Dec of the field centre, along with the GW190814 sky location probability density ($P_\textrm{2D}$) integrated over the field's solid angle. Observations are listed in chronological order per field with the filter and limiting magnitude provided per observation. The exposure time for all observations was 60 seconds. }\label{tab:limmags}\\
\hline
\hline
Field ID & $\int_\textrm{tile}P_\textrm{2D}\,d\Omega$ [\%] & RA [deg] & Dec [deg] & MJD & Filter & Limiting mag [AB] \\
\hline
\endfirsthead
\caption{Continued.} \\
\hline
 Field ID & $\int_\textrm{tile}P_\textrm{2D}\,d\Omega$ [\%] & RA [deg] & Dec [deg] & MJD & Filter & Limiting mag [AB] \\
\hline
\endhead
\hline
\endfoot
\hline
\endlastfoot
3687  & 0.579 & 24.37500 & $-34.12823$ & 58710.116 &\textit{u}& 18.75 \\ 
 & & & & 58710.117 &\textit{q}& 19.79 \\ 
 & & & & 58710.118 &\textit{i}& 19.08 \\ 
 & & & & 58711.115 &\textit{q}& 19.46 \\ 
 & & & & 58711.116 &\textit{i}& 18.85 \\ 
 & & & & 58711.167 &\textit{u}& 18.41 \\ 
 & & & & 58711.168 &\textit{q}& 19.35 \\ 
 & & & & 58711.169 &\textit{i}& 18.49 \\ 
 & & & & 58713.110 &\textit{i}& 18.95 \\ 
 & & & & 58713.161 &\textit{u}& 18.57 \\ 
 & & & & 58713.162 &\textit{q}& 19.47 \\ 
 & & & & 58713.164 &\textit{i}& 18.73 \\ 
 & & & & 58714.137 &\textit{u}& 18.63 \\ 
 & & & & 58714.138 &\textit{q}& 19.77 \\ 
 & & & & 58714.139 &\textit{i}& 18.99 \\ 
 & & & & 58715.096 &\textit{q}& 20.17 \\ 
 & & & & 58715.097 &\textit{i}& 19.41 \\ 
 & & & & 58715.158 &\textit{u}& 18.84 \\ 
 & & & & 58715.160 &\textit{q}& 19.84 \\ 
 & & & & 58715.161 &\textit{i}& 19.17 \\ 
 & & & & 58715.999 &\textit{q}& 20.30 \\ 
 & & & & 58716.000 &\textit{i}& 19.38 \\ 
 & & & & 58716.051 &\textit{u}& 19.20 \\ 
 & & & & 58716.052 &\textit{q}& 20.38 \\ 
 & & & & 58716.053 &\textit{i}& 19.53 \\ 
 & & & & 58716.105 &\textit{u}& 19.16 \\ 
 & & & & 58716.106 &\textit{q}& 20.39 \\ 
 & & & & 58716.107 &\textit{i}& 19.58 \\ 
 & & & & 58717.010 &\textit{q}& 20.18 \\ 
 & & & & 58717.011 &\textit{i}& 19.36 \\ 
 & & & & 58717.064 &\textit{q}& 20.30 \\ 
 & & & & 58717.065 &\textit{i}& 19.42 \\ 
 & & & & 58717.117 &\textit{q}& 20.22 \\ 
 & & & & 58717.118 &\textit{i}& 19.40 \\ 
 & & & & 58717.171 &\textit{q}& 19.76 \\ 
 & & & & 58717.172 &\textit{i}& 18.80 \\ 
 & & & & 58718.097 &\textit{u}& 19.19 \\ 
 & & & & 58718.098 &\textit{q}& 20.36 \\ 
 & & & & 58718.099 &\textit{i}& 19.33 \\ 
 3877 & 3.166 & 21.23077 & $-32.55437$ & 58710.097 &\textit{i}& 19.04 \\ 
 & & & & 58710.160 &\textit{q}& 19.67 \\ 
 & & & & 58710.161 &\textit{i}& 18.90 \\ 
 & & & & 58711.120 &\textit{u}& 18.53 \\ 
 & & & & 58711.121 &\textit{q}& 19.49 \\ 
 & & & & 58711.122 &\textit{i}& 18.86 \\ 
 & & & & 58711.175 &\textit{q}& 19.14 \\ 
 & & & & 58711.176 &\textit{i}& 18.52 \\ 
 & & & & 58713.114 &\textit{u}& 18.70 \\ 
 & & & & 58713.116 &\textit{q}& 19.55 \\ 
 & & & & 58713.117 &\textit{i}& 18.89 \\ 
 & & & & 58713.168 &\textit{u}& 18.52 \\ 
 & & & & 58713.169 &\textit{q}& 19.39 \\ 
 & & & & 58713.170 &\textit{i}& 18.90 \\ 
 & & & & 58714.144 &\textit{u}& 18.72 \\ 
 & & & & 58714.145 &\textit{q}& 19.87 \\ 
 & & & & 58714.146 &\textit{i}& 19.03 \\ 
 & & & & 58715.102 &\textit{u}& 18.93 \\ 
 & & & & 58715.103 &\textit{q}& 20.02 \\ 
 & & & & 58715.104 &\textit{i}& 19.32 \\ 
 & & & & 58715.165 &\textit{u}& 18.91 \\ 
 & & & & 58715.167 &\textit{i}& 18.71 \\ 
 & & & & 58716.004 &\textit{u}& 19.08 \\ 
 & & & & 58716.005 &\textit{q}& 20.23 \\ 
 & & & & 58716.007 &\textit{i}& 19.45 \\ 
 & & & & 58716.058 &\textit{u}& 19.23 \\ 
 & & & & 58716.059 &\textit{q}& 20.28 \\ 
 & & & & 58716.060 &\textit{i}& 19.55 \\ 
 & & & & 58716.112 &\textit{u}& 19.23 \\ 
 & & & & 58716.113 &\textit{q}& 20.38 \\ 
 & & & & 58717.016 &\textit{u}& 19.08 \\ 
 & & & & 58717.017 &\textit{q}& 20.25 \\ 
 & & & & 58717.018 &\textit{i}& 19.39 \\ 
 & & & & 58717.069 &\textit{u}& 19.09 \\ 
 & & & & 58717.070 &\textit{q}& 20.30 \\ 
 & & & & 58717.071 &\textit{i}& 19.49 \\ 
 & & & & 58717.123 &\textit{u}& 19.09 \\ 
 & & & & 58717.124 &\textit{q}& 20.12 \\ 
 & & & & 58717.125 &\textit{i}& 19.26 \\ 
 & & & & 58718.104 &\textit{u}& 19.16 \\ 
 & & & & 58718.105 &\textit{q}& 20.50 \\ 
 & & & & 58718.106 &\textit{i}& 19.43 \\ 
 3878 & 3.682 & 23.07692 & $-32.55437$ & 58709.989 &\textit{i}& 19.04 \\ 
 & & & & 58710.150 &\textit{q}& 19.74 \\ 
 & & & & 58710.151 &\textit{i}& 19.21 \\ 
 & & & & 58711.154 &\textit{u}& 18.60 \\ 
 & & & & 58711.155 &\textit{q}& 19.47 \\ 
 & & & & 58711.156 &\textit{i}& 18.97 \\ 
 & & & & 58713.149 &\textit{q}& 19.49 \\ 
 & & & & 58713.150 &\textit{i}& 18.92 \\ 
 & & & & 58714.124 &\textit{u}& 18.70 \\ 
 & & & & 58714.125 &\textit{q}& 19.78 \\ 
 & & & & 58714.126 &\textit{i}& 18.89 \\ 
 & & & & 58715.083 &\textit{q}& 20.07 \\ 
 & & & & 58715.084 &\textit{i}& 19.43 \\ 
 & & & & 58715.135 &\textit{u}& 19.01 \\ 
 & & & & 58715.136 &\textit{q}& 20.13 \\ 
 & & & & 58715.138 &\textit{i}& 19.46 \\ 
 & & & & 58715.984 &\textit{u}& 18.99 \\ 
 & & & & 58715.985 &\textit{q}& 20.29 \\ 
 & & & & 58715.986 &\textit{i}& 19.46 \\ 
 & & & & 58716.038 &\textit{u}& 19.19 \\ 
 & & & & 58716.039 &\textit{q}& 20.39 \\ 
 & & & & 58716.040 &\textit{i}& 19.45 \\ 
 & & & & 58716.091 &\textit{u}& 19.26 \\ 
 & & & & 58716.093 &\textit{q}& 20.40 \\ 
 & & & & 58716.094 &\textit{i}& 19.57 \\ 
 & & & & 58716.996 &\textit{u}& 18.89 \\ 
 & & & & 58716.997 &\textit{q}& 20.12 \\ 
 & & & & 58716.998 &\textit{i}& 19.24 \\ 
 & & & & 58717.049 &\textit{u}& 19.03 \\ 
 & & & & 58717.050 &\textit{q}& 20.28 \\ 
 & & & & 58717.051 &\textit{i}& 19.46 \\ 
 & & & & 58717.103 &\textit{u}& 19.03 \\ 
 & & & & 58717.104 &\textit{q}& 20.28 \\ 
 & & & & 58717.105 &\textit{i}& 19.43 \\ 
 & & & & 58717.156 &\textit{u}& 18.85 \\ 
 & & & & 58717.157 &\textit{q}& 19.89 \\ 
 & & & & 58717.158 &\textit{i}& 18.97 \\ 
 & & & & 58718.084 &\textit{u}& 19.16 \\ 
 & & & & 58718.085 &\textit{q}& 20.44 \\ 
 & & & & 58718.086 &\textit{i}& 19.44 \\ 
3879 & 0.665 & 24.92308 & $-32.55437$ & 58710.112 &\textit{u}& 18.65 \\ 
 & & & & 58710.113 &\textit{q}& 19.74 \\ 
 & & & & 58710.114 &\textit{i}& 19.15 \\ 
 & & & & 58711.112 &\textit{i}& 18.91 \\ 
 & & & & 58711.164 &\textit{u}& 18.65 \\ 
 & & & & 58711.165 &\textit{q}& 19.54 \\ 
 & & & & 58711.166 &\textit{i}& 18.88 \\ 
 & & & & 58713.158 &\textit{u}& 18.48 \\ 
 & & & & 58713.159 &\textit{q}& 19.48 \\ 
 & & & & 58713.160 &\textit{i}& 18.88 \\ 
 & & & & 58714.134 &\textit{u}& 18.62 \\ 
 & & & & 58714.135 &\textit{q}& 19.76 \\ 
 & & & & 58714.136 &\textit{i}& 18.95 \\ 
 & & & & 58715.093 &\textit{q}& 20.14 \\ 
 & & & & 58715.094 &\textit{i}& 19.40 \\ 
 & & & & 58715.155 &\textit{u}& 19.01 \\ 
 & & & & 58715.156 &\textit{q}& 20.10 \\ 
 & & & & 58715.157 &\textit{i}& 19.23 \\ 
 & & & & 58715.994 &\textit{u}& 19.11 \\ 
 & & & & 58715.995 &\textit{q}& 20.26 \\ 
 & & & & 58715.997 &\textit{i}& 19.45 \\ 
 & & & & 58716.048 &\textit{u}& 19.20 \\ 
 & & & & 58716.049 &\textit{q}& 20.30 \\ 
 & & & & 58716.050 &\textit{i}& 19.57 \\ 
 & & & & 58716.102 &\textit{u}& 19.19 \\ 
 & & & & 58716.103 &\textit{q}& 20.35 \\ 
 & & & & 58716.104 &\textit{i}& 19.63 \\ 
 & & & & 58717.007 &\textit{q}& 20.17 \\ 
 & & & & 58717.008 &\textit{i}& 19.34 \\ 
 & & & & 58717.060 &\textit{q}& 20.27 \\ 
 & & & & 58717.062 &\textit{i}& 19.46 \\ 
 & & & & 58717.114 &\textit{q}& 20.27 \\ 
 & & & & 58717.115 &\textit{i}& 19.43 \\ 
 & & & & 58717.166 &\textit{u}& 18.72 \\ 
 & & & & 58717.168 &\textit{q}& 19.83 \\ 
 & & & & 58717.169 &\textit{i}& 18.86 \\ 
 & & & & 58718.094 &\textit{u}& 19.20 \\ 
 & & & & 58718.095 &\textit{q}& 20.45 \\ 
 & & & & 58718.096 &\textit{i}& 19.33 \\ 
4073 & 2.441 & 21.81818 & $-30.98052$ & 58710.156 &\textit{u}& 18.66 \\ 
 & & & & 58710.157 &\textit{q}& 19.70 \\ 
 & & & & 58710.158 &\textit{i}& 19.02 \\ 
 & & & & 58711.117 &\textit{u}& 18.61 \\ 
 & & & & 58711.118 &\textit{q}& 19.38 \\ 
 & & & & 58711.119 &\textit{i}& 18.92 \\ 
 & & & & 58711.170 &\textit{u}& 18.34 \\ 
 & & & & 58711.172 &\textit{q}& 19.36 \\ 
 & & & & 58711.173 &\textit{i}& 18.72 \\ 
 & & & & 58713.111 &\textit{u}& 18.59 \\ 
 & & & & 58713.112 &\textit{q}& 19.57 \\ 
 & & & & 58713.113 &\textit{i}& 18.98 \\ 
 & & & & 58713.167 &\textit{i}& 18.69 \\ 
 & & & & 58714.141 &\textit{u}& 18.75 \\ 
 & & & & 58714.142 &\textit{q}& 19.78 \\ 
 & & & & 58714.143 &\textit{i}& 19.02 \\ 
 & & & & 58715.098 &\textit{u}& 19.00 \\ 
 & & & & 58715.099 &\textit{q}& 19.97 \\ 
 & & & & 58715.101 &\textit{i}& 19.19 \\ 
 & & & & 58715.162 &\textit{u}& 18.92 \\ 
 & & & & 58715.164 &\textit{i}& 18.66 \\ 
 & & & & 58716.001 &\textit{u}& 19.06 \\ 
 & & & & 58716.002 &\textit{q}& 20.20 \\ 
 & & & & 58716.003 &\textit{i}& 19.44 \\ 
 & & & & 58716.055 &\textit{u}& 19.21 \\ 
 & & & & 58716.056 &\textit{q}& 20.37 \\ 
 & & & & 58716.057 &\textit{i}& 19.50 \\ 
 & & & & 58716.108 &\textit{u}& 19.18 \\ 
 & & & & 58716.109 &\textit{q}& 20.29 \\ 
 & & & & 58716.976 &\textit{u}& 18.98 \\ 
 & & & & 58717.013 &\textit{u}& 19.07 \\ 
 & & & & 58717.014 &\textit{q}& 20.17 \\ 
 & & & & 58717.015 &\textit{i}& 19.41 \\ 
 & & & & 58717.066 &\textit{u}& 19.02 \\ 
 & & & & 58717.067 &\textit{q}& 20.21 \\ 
 & & & & 58717.068 &\textit{i}& 19.41 \\ 
 & & & & 58717.119 &\textit{u}& 19.06 \\ 
 & & & & 58717.121 &\textit{q}& 20.14 \\ 
 & & & & 58717.122 &\textit{i}& 19.19 \\ 
 & & & & 58717.174 &\textit{q}& 19.57 \\ 
 & & & & 58717.175 &\textit{i}& 18.72 \\ 
 & & & & 58718.100 &\textit{u}& 19.15 \\ 
 & & & & 58718.102 &\textit{q}& 20.39 \\ 
 & & & & 58718.103 &\textit{i}& 19.41 \\ 
4467 & 0.320 & 12.35294 & $-27.83282$ & 58710.109 &\textit{u}& 18.56 \\ 
 & & & & 58710.110 &\textit{q}& 19.66 \\ 
 & & & & 58710.111 &\textit{i}& 18.95 \\ 
 & & & & 58711.107 &\textit{u}& 18.48 \\ 
 & & & & 58711.108 &\textit{q}& 19.22 \\ 
 & & & & 58711.109 &\textit{i}& 18.77 \\ 
 & & & & 58711.161 &\textit{q}& 19.14 \\ 
 & & & & 58711.162 &\textit{i}& 18.55 \\ 
 & & & & 58713.101 &\textit{u}& 18.53 \\ 
 & & & & 58713.102 &\textit{q}& 19.35 \\ 
 & & & & 58713.103 &\textit{i}& 18.79 \\ 
 & & & & 58714.130 &\textit{u}& 18.66 \\ 
 & & & & 58714.132 &\textit{q}& 19.57 \\ 
 & & & & 58714.133 &\textit{i}& 18.71 \\ 
 & & & & 58715.037 &\textit{i}& 18.81 \\ 
 & & & & 58715.088 &\textit{u}& 19.00 \\ 
 & & & & 58715.089 &\textit{q}& 19.84 \\ 
 & & & & 58715.091 &\textit{i}& 18.88 \\ 
 & & & & 58715.142 &\textit{u}& 18.84 \\ 
 & & & & 58715.144 &\textit{i}& 18.88 \\ 
 & & & & 58715.991 &\textit{u}& 19.06 \\ 
 & & & & 58715.992 &\textit{q}& 20.19 \\ 
 & & & & 58716.044 &\textit{u}& 19.19 \\ 
 & & & & 58716.046 &\textit{q}& 20.20 \\ 
 & & & & 58716.047 &\textit{i}& 19.12 \\ 
 & & & & 58716.098 &\textit{u}& 19.14 \\ 
 & & & & 58716.099 &\textit{q}& 20.22 \\ 
 & & & & 58716.100 &\textit{i}& 19.10 \\ 
 & & & & 58717.003 &\textit{u}& 19.04 \\ 
 & & & & 58717.004 &\textit{q}& 20.14 \\ 
 & & & & 58717.005 &\textit{i}& 19.39 \\ 
 & & & & 58717.056 &\textit{u}& 19.05 \\ 
 & & & & 58717.057 &\textit{q}& 20.08 \\ 
 & & & & 58717.058 &\textit{i}& 19.22 \\ 
 & & & & 58717.109 &\textit{u}& 18.96 \\ 
 & & & & 58717.110 &\textit{q}& 20.08 \\ 
 & & & & 58717.112 &\textit{i}& 19.09 \\ 
 & & & & 58717.164 &\textit{q}& 19.87 \\ 
 & & & & 58717.165 &\textit{i}& 18.95 \\ 
 & & & & 58718.090 &\textit{u}& 19.16 \\ 
 & & & & 58718.091 &\textit{q}& 20.47 \\ 
 & & & & 58718.093 &\textit{i}& 19.47 \\ 
4468 & 0.768 & 14.11765 & $-27.83282$ & 58710.085 &\textit{u}& 18.64 \\ 
 & & & & 58710.086 &\textit{q}& 19.72 \\ 
 & & & & 58710.087 &\textit{i}& 19.06 \\ 
 & & & & 58710.166 &\textit{u}& 18.75 \\ 
 & & & & 58710.167 &\textit{q}& 19.77 \\ 
 & & & & 58710.168 &\textit{i}& 19.14 \\ 
 & & & & 58711.103 &\textit{u}& 18.41 \\ 
 & & & & 58711.104 &\textit{q}& 19.36 \\ 
 & & & & 58711.106 &\textit{i}& 18.85 \\ 
 & & & & 58711.158 &\textit{q}& 19.11 \\ 
 & & & & 58711.159 &\textit{i}& 18.58 \\ 
 & & & & 58713.099 &\textit{q}& 19.42 \\ 
 & & & & 58713.100 &\textit{i}& 18.82 \\ 
 & & & & 58713.152 &\textit{q}& 19.34 \\ 
 & & & & 58713.153 &\textit{i}& 18.75 \\ 
 & & & & 58714.128 &\textit{q}& 19.74 \\ 
 & & & & 58714.129 &\textit{i}& 18.90 \\ 
 & & & & 58715.085 &\textit{u}& 19.01 \\ 
 & & & & 58715.086 &\textit{q}& 19.85 \\ 
 & & & & 58715.087 &\textit{i}& 18.91 \\ 
 & & & & 58715.139 &\textit{u}& 18.96 \\ 
 & & & & 58715.140 &\textit{q}& 19.87 \\ 
 & & & & 58715.141 &\textit{i}& 18.96 \\ 
 & & & & 58715.989 &\textit{q}& 20.21 \\ 
 & & & & 58715.990 &\textit{i}& 19.30 \\ 
 & & & & 58716.041 &\textit{u}& 19.19 \\ 
 & & & & 58716.042 &\textit{q}& 20.18 \\ 
 & & & & 58716.043 &\textit{i}& 19.03 \\ 
 & & & & 58716.095 &\textit{u}& 19.13 \\ 
 & & & & 58716.096 &\textit{q}& 20.20 \\ 
 & & & & 58716.097 &\textit{i}& 19.23 \\ 
 & & & & 58717.000 &\textit{q}& 20.16 \\ 
 & & & & 58717.001 &\textit{i}& 19.38 \\ 
 & & & & 58717.053 &\textit{u}& 19.01 \\ 
 & & & & 58717.054 &\textit{q}& 20.15 \\ 
 & & & & 58717.055 &\textit{i}& 19.19 \\ 
 & & & & 58717.106 &\textit{u}& 18.95 \\ 
 & & & & 58717.107 &\textit{q}& 20.11 \\ 
 & & & & 58717.108 &\textit{i}& 19.10 \\ 
 & & & & 58717.161 &\textit{q}& 19.87 \\ 
 & & & & 58717.162 &\textit{i}& 19.05 \\ 
 & & & & 58718.087 &\textit{u}& 19.14 \\ 
 & & & & 58718.088 &\textit{q}& 20.44 \\ 
 & & & & 58718.089 &\textit{i}& 19.45 \\ 
4670 & 3.858 & 11.30435 & $-26.25896$ & 58710.072 &\textit{u}& 18.52 \\ 
 & & & & 58710.073 &\textit{q}& 19.75 \\ 
 & & & & 58710.074 &\textit{i}& 19.05 \\ 
 & & & & 58710.152 &\textit{u}& 18.57 \\ 
 & & & & 58710.153 &\textit{q}& 19.80 \\ 
 & & & & 58710.155 &\textit{i}& 19.13 \\ 
 & & & & 58711.131 &\textit{q}& 19.11 \\ 
 & & & & 58711.132 &\textit{i}& 18.59 \\ 
 & & & & 58713.125 &\textit{u}& 18.52 \\ 
 & & & & 58713.126 &\textit{q}& 19.23 \\ 
 & & & & 58713.127 &\textit{i}& 18.83 \\ 
 & & & & 58714.951 &\textit{u}& 18.84 \\ 
 & & & & 58714.952 &\textit{q}& 19.95 \\ 
 & & & & 58714.953 &\textit{i}& 19.26 \\ 
 & & & & 58715.059 &\textit{q}& 19.96 \\ 
 & & & & 58715.060 &\textit{i}& 19.08 \\ 
 & & & & 58715.112 &\textit{u}& 18.90 \\ 
 & & & & 58715.113 &\textit{q}& 19.92 \\ 
 & & & & 58715.114 &\textit{i}& 18.96 \\ 
 & & & & 58715.175 &\textit{u}& 18.60 \\ 
 & & & & 58715.176 &\textit{q}& 19.60 \\ 
 & & & & 58715.177 &\textit{i}& 18.77 \\ 
 & & & & 58715.961 &\textit{u}& 19.06 \\ 
 & & & & 58715.962 &\textit{q}& 20.22 \\ 
 & & & & 58715.963 &\textit{i}& 19.43 \\ 
 & & & & 58716.014 &\textit{u}& 19.16 \\ 
 & & & & 58716.016 &\textit{q}& 19.99 \\ 
 & & & & 58716.017 &\textit{i}& 19.12 \\ 
 & & & & 58716.068 &\textit{u}& 19.23 \\ 
 & & & & 58716.069 &\textit{q}& 20.20 \\ 
 & & & & 58716.070 &\textit{i}& 19.25 \\ 
 & & & & 58716.972 &\textit{u}& 19.01 \\ 
 & & & & 58716.974 &\textit{q}& 20.10 \\ 
 & & & & 58716.975 &\textit{i}& 19.18 \\ 
 & & & & 58717.019 &\textit{u}& 19.02 \\ 
 & & & & 58717.027 &\textit{q}& 20.07 \\ 
 & & & & 58717.028 &\textit{i}& 19.16 \\ 
 & & & & 58717.079 &\textit{u}& 19.06 \\ 
 & & & & 58717.080 &\textit{q}& 20.04 \\ 
 & & & & 58717.082 &\textit{i}& 19.13 \\ 
 & & & & 58717.133 &\textit{u}& 18.81 \\ 
 & & & & 58717.134 &\textit{q}& 19.96 \\ 
 & & & & 58717.135 &\textit{i}& 19.22 \\ 
 & & & & 58718.070 &\textit{u}& 19.19 \\ 
 & & & & 58718.071 &\textit{q}& 20.46 \\ 
 & & & & 58718.072 &\textit{i}& 19.41 \\ 
4671 & 17.861 & 13.04348 & $-26.25896$ & 58710.137 &\textit{q}& 19.47 \\ 
 & & & & 58710.138 &\textit{i}& 18.77 \\ 
 & & & & 58711.137 &\textit{u}& 18.37 \\ 
 & & & & 58711.138 &\textit{q}& 19.14 \\ 
 & & & & 58711.139 &\textit{i}& 18.57 \\ 
 & & & & 58713.131 &\textit{u}& 18.52 \\ 
 & & & & 58713.132 &\textit{q}& 19.30 \\ 
 & & & & 58713.133 &\textit{i}& 18.83 \\ 
 & & & & 58714.161 &\textit{u}& 18.66 \\ 
 & & & & 58714.957 &\textit{u}& 18.92 \\ 
 & & & & 58714.958 &\textit{q}& 20.01 \\ 
 & & & & 58714.959 &\textit{i}& 19.26 \\ 
 & & & & 58715.066 &\textit{q}& 19.99 \\ 
 & & & & 58715.067 &\textit{i}& 19.05 \\ 
 & & & & 58715.119 &\textit{u}& 18.75 \\ 
 & & & & 58715.120 &\textit{q}& 19.91 \\ 
 & & & & 58715.121 &\textit{i}& 19.08 \\ 
 & & & & 58715.967 &\textit{u}& 19.07 \\ 
 & & & & 58715.969 &\textit{q}& 20.16 \\ 
 & & & & 58715.970 &\textit{i}& 19.43 \\ 
 & & & & 58716.021 &\textit{u}& 19.12 \\ 
 & & & & 58716.022 &\textit{q}& 20.10 \\ 
 & & & & 58716.023 &\textit{i}& 19.10 \\ 
 & & & & 58716.075 &\textit{u}& 19.21 \\ 
 & & & & 58716.076 &\textit{q}& 20.16 \\ 
 & & & & 58716.077 &\textit{i}& 19.18 \\ 
 & & & & 58716.979 &\textit{u}& 18.96 \\ 
 & & & & 58716.980 &\textit{q}& 20.10 \\ 
 & & & & 58716.981 &\textit{i}& 19.24 \\ 
 & & & & 58717.033 &\textit{u}& 19.05 \\ 
 & & & & 58717.034 &\textit{q}& 20.04 \\ 
 & & & & 58717.035 &\textit{i}& 19.17 \\ 
 & & & & 58717.087 &\textit{q}& 20.07 \\ 
 & & & & 58717.088 &\textit{i}& 19.16 \\ 
 & & & & 58717.141 &\textit{q}& 20.02 \\ 
 & & & & 58717.142 &\textit{i}& 19.14 \\ 
 & & & & 58718.060 &\textit{u}& 19.12 \\ 
 & & & & 58718.061 &\textit{q}& 20.39 \\ 
 & & & & 58718.062 &\textit{i}& 19.44 \\ 
4672 & 3.763 & 14.78261 & $-26.25896$ & 58710.146 &\textit{u}& 18.50 \\ 
 & & & & 58710.147 &\textit{q}& 19.60 \\ 
 & & & & 58710.148 &\textit{i}& 19.09 \\ 
 & & & & 58711.099 &\textit{i}& 18.80 \\ 
 & & & & 58711.150 &\textit{u}& 18.32 \\ 
 & & & & 58711.151 &\textit{q}& 19.21 \\ 
 & & & & 58711.152 &\textit{i}& 18.66 \\ 
 & & & & 58713.144 &\textit{u}& 18.58 \\ 
 & & & & 58713.146 &\textit{q}& 19.35 \\ 
 & & & & 58713.147 &\textit{i}& 18.80 \\ 
 & & & & 58714.120 &\textit{u}& 18.71 \\ 
 & & & & 58714.121 &\textit{q}& 19.78 \\ 
 & & & & 58714.122 &\textit{i}& 18.94 \\ 
 & & & & 58714.972 &\textit{q}& 19.92 \\ 
 & & & & 58715.079 &\textit{q}& 19.99 \\ 
 & & & & 58715.080 &\textit{i}& 19.16 \\ 
 & & & & 58715.133 &\textit{q}& 19.85 \\ 
 & & & & 58715.134 &\textit{i}& 19.07 \\ 
 & & & & 58715.981 &\textit{u}& 19.06 \\ 
 & & & & 58715.982 &\textit{q}& 20.11 \\ 
 & & & & 58715.983 &\textit{i}& 19.07 \\ 
 & & & & 58716.034 &\textit{u}& 19.08 \\ 
 & & & & 58716.035 &\textit{q}& 20.14 \\ 
 & & & & 58716.037 &\textit{i}& 19.06 \\ 
 & & & & 58716.088 &\textit{u}& 19.23 \\ 
 & & & & 58716.089 &\textit{q}& 20.05 \\ 
 & & & & 58716.090 &\textit{i}& 19.10 \\ 
 & & & & 58716.992 &\textit{u}& 18.99 \\ 
 & & & & 58716.994 &\textit{q}& 20.14 \\ 
 & & & & 58716.995 &\textit{i}& 19.03 \\ 
 & & & & 58717.046 &\textit{u}& 19.07 \\ 
 & & & & 58717.047 &\textit{q}& 20.10 \\ 
 & & & & 58717.048 &\textit{i}& 19.17 \\ 
 & & & & 58717.099 &\textit{u}& 19.06 \\ 
 & & & & 58717.100 &\textit{q}& 20.16 \\ 
 & & & & 58717.101 &\textit{i}& 19.20 \\ 
 & & & & 58717.154 &\textit{q}& 19.91 \\ 
 & & & & 58717.155 &\textit{i}& 18.97 \\ 
 & & & & 58718.077 &\textit{u}& 19.10 \\ 
 & & & & 58718.078 &\textit{q}& 20.42 \\ 
 & & & & 58718.079 &\textit{i}& 19.41 \\ 
4877 & 2.190 & 10.28571 & $-24.68511$ & 58710.082 &\textit{u}& 18.59 \\ 
 & & & & 58710.083 &\textit{q}& 19.69 \\ 
 & & & & 58710.084 &\textit{i}& 19.08 \\ 
 & & & & 58710.163 &\textit{u}& 18.70 \\ 
 & & & & 58710.164 &\textit{q}& 19.77 \\ 
 & & & & 58710.165 &\textit{i}& 19.03 \\ 
 & & & & 58711.128 &\textit{q}& 19.14 \\ 
 & & & & 58711.129 &\textit{i}& 18.63 \\ 
 & & & & 58713.121 &\textit{u}& 18.48 \\ 
 & & & & 58713.122 &\textit{q}& 19.32 \\ 
 & & & & 58713.123 &\textit{i}& 18.81 \\ 
 & & & & 58714.947 &\textit{u}& 18.89 \\ 
 & & & & 58714.948 &\textit{q}& 19.97 \\ 
 & & & & 58714.949 &\textit{i}& 19.30 \\ 
 & & & & 58715.055 &\textit{u}& 18.97 \\ 
 & & & & 58715.056 &\textit{q}& 19.86 \\ 
 & & & & 58715.057 &\textit{i}& 19.06 \\ 
 & & & & 58715.109 &\textit{u}& 18.89 \\ 
 & & & & 58715.110 &\textit{q}& 19.84 \\ 
 & & & & 58715.172 &\textit{u}& 18.79 \\ 
 & & & & 58715.173 &\textit{q}& 19.76 \\ 
 & & & & 58715.174 &\textit{i}& 18.96 \\ 
 & & & & 58715.958 &\textit{u}& 19.04 \\ 
 & & & & 58715.959 &\textit{q}& 20.23 \\ 
 & & & & 58715.960 &\textit{i}& 19.39 \\ 
 & & & & 58716.011 &\textit{u}& 19.05 \\ 
 & & & & 58716.012 &\textit{q}& 20.10 \\ 
 & & & & 58716.013 &\textit{i}& 19.07 \\ 
 & & & & 58716.065 &\textit{u}& 19.12 \\ 
 & & & & 58716.066 &\textit{q}& 20.13 \\ 
 & & & & 58716.067 &\textit{i}& 19.17 \\ 
 & & & & 58716.969 &\textit{u}& 19.02 \\ 
 & & & & 58716.970 &\textit{q}& 20.05 \\ 
 & & & & 58716.971 &\textit{i}& 19.18 \\ 
 & & & & 58717.023 &\textit{u}& 18.98 \\ 
 & & & & 58717.024 &\textit{q}& 20.04 \\ 
 & & & & 58717.025 &\textit{i}& 19.13 \\ 
 & & & & 58717.076 &\textit{u}& 19.01 \\ 
 & & & & 58717.077 &\textit{q}& 20.09 \\ 
 & & & & 58717.078 &\textit{i}& 19.06 \\ 
 & & & & 58717.130 &\textit{u}& 18.81 \\ 
 & & & & 58717.131 &\textit{q}& 20.06 \\ 
 & & & & 58717.132 &\textit{i}& 19.22 \\ 
 & & & & 58718.080 &\textit{u}& 19.15 \\ 
 & & & & 58718.081 &\textit{q}& 20.44 \\ 
 & & & & 58718.082 &\textit{i}& 19.57 \\ 
4878 & 31.618 & 12.00000 & $-24.68511$ & 58710.130 &\textit{q}& 19.46 \\ 
 & & & & 58710.131 &\textit{i}& 18.63 \\ 
 & & & & 58711.134 &\textit{u}& 18.28 \\ 
 & & & & 58711.135 &\textit{q}& 19.11 \\ 
 & & & & 58711.136 &\textit{i}& 18.61 \\ 
 & & & & 58713.128 &\textit{u}& 18.46 \\ 
 & & & & 58713.129 &\textit{q}& 19.23 \\ 
 & & & & 58713.130 &\textit{i}& 18.80 \\ 
 & & & & 58714.954 &\textit{u}& 18.83 \\ 
 & & & & 58714.955 &\textit{q}& 20.00 \\ 
 & & & & 58714.956 &\textit{i}& 19.26 \\ 
 & & & & 58715.062 &\textit{q}& 19.87 \\ 
 & & & & 58715.063 &\textit{i}& 19.09 \\ 
 & & & & 58715.115 &\textit{u}& 18.86 \\ 
 & & & & 58715.116 &\textit{q}& 19.86 \\ 
 & & & & 58715.964 &\textit{u}& 18.98 \\ 
 & & & & 58715.965 &\textit{q}& 20.19 \\ 
 & & & & 58715.966 &\textit{i}& 19.43 \\ 
 & & & & 58716.018 &\textit{u}& 19.01 \\ 
 & & & & 58716.019 &\textit{q}& 20.07 \\ 
 & & & & 58716.020 &\textit{i}& 19.05 \\ 
 & & & & 58716.071 &\textit{u}& 19.22 \\ 
 & & & & 58716.073 &\textit{q}& 20.19 \\ 
 & & & & 58716.074 &\textit{i}& 19.17 \\ 
 & & & & 58716.976 &\textit{u}& 18.98 \\ 
 & & & & 58716.977 &\textit{q}& 20.16 \\ 
 & & & & 58716.978 &\textit{i}& 19.25 \\ 
 & & & & 58717.029 &\textit{u}& 19.01 \\ 
 & & & & 58717.030 &\textit{q}& 20.10 \\ 
 & & & & 58717.031 &\textit{i}& 19.15 \\ 
 & & & & 58717.083 &\textit{u}& 18.93 \\ 
 & & & & 58717.084 &\textit{q}& 20.11 \\ 
 & & & & 58717.085 &\textit{i}& 19.15 \\ 
 & & & & 58717.137 &\textit{q}& 20.03 \\ 
 & & & & 58717.138 &\textit{i}& 19.13 \\ 
 & & & & 58718.057 &\textit{u}& 19.17 \\ 
 & & & & 58718.058 &\textit{q}& 20.46 \\ 
4879 & 19.693 & 13.71429 & $-24.68511$ & 58710.133 &\textit{q}& 19.36 \\ 
 & & & & 58710.134 &\textit{i}& 18.69 \\ 
 & & & & 58711.141 &\textit{q}& 19.20 \\ 
 & & & & 58711.142 &\textit{i}& 18.66 \\ 
 & & & & 58713.136 &\textit{q}& 19.22 \\ 
 & & & & 58714.961 &\textit{u}& 18.95 \\ 
 & & & & 58714.962 &\textit{q}& 19.98 \\ 
 & & & & 58714.963 &\textit{i}& 19.28 \\ 
 & & & & 58715.068 &\textit{u}& 18.98 \\ 
 & & & & 58715.069 &\textit{q}& 19.89 \\ 
 & & & & 58715.070 &\textit{i}& 19.06 \\ 
 & & & & 58715.123 &\textit{q}& 19.79 \\ 
 & & & & 58715.124 &\textit{i}& 19.09 \\ 
 & & & & 58715.972 &\textit{q}& 20.13 \\ 
 & & & & 58715.973 &\textit{i}& 19.39 \\ 
 & & & & 58716.024 &\textit{u}& 19.13 \\ 
 & & & & 58716.025 &\textit{q}& 20.13 \\ 
 & & & & 58716.027 &\textit{i}& 19.01 \\ 
 & & & & 58716.078 &\textit{u}& 19.05 \\ 
 & & & & 58716.079 &\textit{q}& 20.17 \\ 
 & & & & 58716.080 &\textit{i}& 19.13 \\ 
 & & & & 58716.982 &\textit{u}& 19.00 \\ 
 & & & & 58716.984 &\textit{q}& 20.10 \\ 
 & & & & 58716.985 &\textit{i}& 19.32 \\ 
 & & & & 58717.036 &\textit{u}& 18.96 \\ 
 & & & & 58717.037 &\textit{q}& 20.06 \\ 
 & & & & 58717.038 &\textit{i}& 19.14 \\ 
 & & & & 58717.089 &\textit{u}& 19.04 \\ 
 & & & & 58717.090 &\textit{q}& 20.11 \\ 
 & & & & 58717.091 &\textit{i}& 19.09 \\ 
 & & & & 58717.144 &\textit{q}& 20.01 \\ 
 & & & & 58717.145 &\textit{i}& 19.19 \\ 
 & & & & 58718.064 &\textit{u}& 19.21 \\ 
 & & & & 58718.065 &\textit{q}& 20.38 \\ 
 & & & & 58718.066 &\textit{i}& 19.40 \\ 
5087 & 6.396 & 11.03773 & $-23.11126$ & 58710.142 &\textit{u}& 18.47 \\ 
 & & & & 58710.143 &\textit{q}& 19.54 \\ 
 & & & & 58710.144 &\textit{i}& 18.97 \\ 
 & & & & 58711.094 &\textit{q}& 19.33 \\ 
 & & & & 58711.095 &\textit{i}& 18.77 \\ 
 & & & & 58711.148 &\textit{q}& 18.99 \\ 
 & & & & 58711.149 &\textit{i}& 18.57 \\ 
 & & & & 58714.968 &\textit{q}& 19.86 \\ 
 & & & & 58714.969 &\textit{i}& 18.86 \\ 
 & & & & 58715.075 &\textit{u}& 18.85 \\ 
 & & & & 58715.076 &\textit{q}& 19.80 \\ 
 & & & & 58715.077 &\textit{i}& 18.95 \\ 
 & & & & 58715.129 &\textit{u}& 18.92 \\ 
 & & & & 58715.130 &\textit{q}& 19.82 \\ 
 & & & & 58715.131 &\textit{i}& 18.94 \\ 
 & & & & 58715.977 &\textit{u}& 19.00 \\ 
 & & & & 58715.978 &\textit{q}& 20.04 \\ 
 & & & & 58715.980 &\textit{i}& 19.09 \\ 
 & & & & 58716.031 &\textit{u}& 19.09 \\ 
 & & & & 58716.032 &\textit{q}& 20.10 \\ 
 & & & & 58716.033 &\textit{i}& 19.12 \\ 
 & & & & 58716.085 &\textit{u}& 19.14 \\ 
 & & & & 58716.086 &\textit{q}& 20.20 \\ 
 & & & & 58716.087 &\textit{i}& 19.10 \\ 
 & & & & 58716.989 &\textit{u}& 18.98 \\ 
 & & & & 58716.990 &\textit{q}& 20.09 \\ 
 & & & & 58716.991 &\textit{i}& 19.12 \\ 
 & & & & 58717.042 &\textit{u}& 18.96 \\ 
 & & & & 58717.044 &\textit{q}& 20.12 \\ 
 & & & & 58717.045 &\textit{i}& 19.15 \\ 
 & & & & 58717.096 &\textit{u}& 18.91 \\ 
 & & & & 58717.097 &\textit{q}& 20.09 \\ 
 & & & & 58717.098 &\textit{i}& 19.06 \\ 
 & & & & 58717.149 &\textit{u}& 18.70 \\ 
 & & & & 58717.151 &\textit{q}& 19.84 \\ 
 & & & & 58717.152 &\textit{i}& 18.97 \\ 
 & & & & 58718.074 &\textit{u}& 19.07 \\ 
 & & & & 58718.075 &\textit{q}& 20.43 \\ 
 & & & & 58718.076 &\textit{i}& 19.47 \\ 
5088 &7.722 & 12.73585 & $-23.11126$ & 58710.139 &\textit{u}& 18.46 \\ 
 & & & & 58710.140 &\textit{q}& 19.52 \\ 
 & & & & 58710.141 &\textit{i}& 18.74 \\ 
 & & & & 58711.092 &\textit{i}& 18.85 \\ 
 & & & & 58711.145 &\textit{q}& 19.10 \\ 
 & & & & 58711.146 &\textit{i}& 18.68 \\ 
 & & & & 58713.138 &\textit{u}& 18.44 \\ 
 & & & & 58713.139 &\textit{q}& 19.29 \\ 
 & & & & 58713.140 &\textit{i}& 18.67 \\ 
 & & & & 58714.964 &\textit{u}& 18.90 \\ 
 & & & & 58714.965 &\textit{q}& 19.98 \\ 
 & & & & 58714.966 &\textit{i}& 19.28 \\ 
 & & & & 58715.071 &\textit{u}& 18.98 \\ 
 & & & & 58715.072 &\textit{q}& 19.75 \\ 
 & & & & 58715.073 &\textit{i}& 18.87 \\ 
 & & & & 58715.125 &\textit{u}& 18.97 \\ 
 & & & & 58715.126 &\textit{q}& 19.80 \\ 
 & & & & 58715.127 &\textit{i}& 18.84 \\ 
 & & & & 58715.974 &\textit{u}& 19.01 \\ 
 & & & & 58715.975 &\textit{q}& 20.14 \\ 
 & & & & 58715.976 &\textit{i}& 19.36 \\ 
 & & & & 58716.028 &\textit{u}& 19.13 \\ 
 & & & & 58716.029 &\textit{q}& 20.09 \\ 
 & & & & 58716.081 &\textit{u}& 19.23 \\ 
 & & & & 58716.082 &\textit{q}& 20.16 \\ 
 & & & & 58716.986 &\textit{u}& 19.05 \\ 
 & & & & 58716.987 &\textit{q}& 20.05 \\ 
 & & & & 58716.988 &\textit{i}& 19.08 \\ 
 & & & & 58717.039 &\textit{u}& 19.03 \\ 
 & & & & 58717.040 &\textit{q}& 20.07 \\ 
 & & & & 58717.041 &\textit{i}& 19.12 \\ 
 & & & & 58717.093 &\textit{u}& 18.99 \\ 
 & & & & 58717.094 &\textit{q}& 20.09 \\ 
 & & & & 58717.095 &\textit{i}& 19.15 \\ 
 & & & & 58717.146 &\textit{u}& 18.83 \\ 
 & & & & 58717.147 &\textit{q}& 19.98 \\ 
 & & & & 58717.148 &\textit{i}& 19.13 \\ 
 & & & & 58718.067 &\textit{u}& 19.24 \\ 
 & & & & 58718.068 &\textit{q}& 20.47 \\ 
 & & & & 58718.069 &\textit{i}& 19.51 \\ 
5089 & 0.418 & 14.43396 & $-23.11126$ & 58710.095 &\textit{u}& 18.65 \\ 
 & & & & 58710.096 &\textit{q}& 19.59 \\ 
 & & & & 58710.097 &\textit{i}& 19.04 \\ 
 & & & & 58710.176 &\textit{u}& 18.78 \\ 
 & & & & 58710.177 &\textit{q}& 19.81 \\ 
 & & & & 58710.178 &\textit{i}& 19.00 \\ 
 & & & & 58711.124 &\textit{u}& 18.43 \\ 
 & & & & 58711.125 &\textit{q}& 19.32 \\ 
 & & & & 58711.126 &\textit{i}& 18.68 \\ 
 & & & & 58711.177 &\textit{u}& 18.19 \\ 
 & & & & 58711.178 &\textit{q}& 18.88 \\ 
 & & & & 58711.179 &\textit{i}& 18.25 \\ 
 & & & & 58713.118 &\textit{u}& 18.50 \\ 
 & & & & 58713.119 &\textit{q}& 19.41 \\ 
 & & & & 58713.120 &\textit{i}& 18.75 \\ 
 & & & & 58715.053 &\textit{i}& 18.82 \\ 
 & & & & 58715.105 &\textit{u}& 18.90 \\ 
 & & & & 58715.106 &\textit{q}& 19.78 \\ 
 & & & & 58715.107 &\textit{i}& 18.88 \\ 
 & & & & 58715.168 &\textit{u}& 18.77 \\ 
 & & & & 58715.170 &\textit{q}& 19.78 \\ 
 & & & & 58715.171 &\textit{i}& 18.84 \\ 
 & & & & 58716.008 &\textit{u}& 19.14 \\ 
 & & & & 58716.009 &\textit{q}& 20.05 \\ 
 & & & & 58716.010 &\textit{i}& 19.02 \\ 
 & & & & 58716.061 &\textit{u}& 19.16 \\ 
 & & & & 58716.062 &\textit{q}& 20.13 \\ 
 & & & & 58716.064 &\textit{i}& 19.17 \\ 
 & & & & 58716.115 &\textit{u}& 19.24 \\ 
 & & & & 58716.116 &\textit{q}& 20.07 \\ 
 & & & & 58716.117 &\textit{i}& 19.12 \\ 
 & & & & 58717.019 &\textit{u}& 19.02 \\ 
 & & & & 58717.020 &\textit{q}& 20.05 \\ 
 & & & & 58717.021 &\textit{i}& 19.02 \\ 
 & & & & 58717.073 &\textit{u}& 19.10 \\ 
 & & & & 58717.074 &\textit{q}& 20.01 \\ 
 & & & & 58717.075 &\textit{i}& 19.08 \\ 
 & & & & 58717.127 &\textit{q}& 20.05 \\ 
 & & & & 58717.128 &\textit{i}& 19.16 \\ 
\end{longtable}
}

\end{appendix}

\end{document}